\documentclass[reprint,amsmath,amssymb,aps]{revtex4-2}
\usepackage[latin9]{inputenc}
\usepackage{amsmath}
\usepackage{amsthm}
\usepackage{amssymb}
\usepackage{graphicx}
\usepackage{placeins}

\makeatletter
\usepackage{dcolumn}
\usepackage{bm}
\usepackage{braket}
\usepackage{amsthm}

\makeatother

\begin{document}
\title{Fractional Charges, Linear Conditions and Chemical Potentials for
Excited States in $\Delta SCF$ Theory}
\author{Weitao Yang}
\affiliation{Department of Chemistry and Department of Physics, Duke University,
Durham, North Carolina 27708}
\author{Yichen Fan}
\affiliation{Department of Chemistry, Duke University, Durham, North Carolina 27708}
\begin{abstract}
To describe excited states, the electron density alone is insufficient.
Instead, we use the noninteracting reference density matrix $\gamma_{s}({\bf x},{\bf x}')$
based on the recently established foundation for the $\Delta$SCF theory, in which ground and excited state energies and densities are
obtained from the minimum and stationary solutions of the same functional.
We now extend the theory to fractional charges. Based on the exact
properties of degeneracy and size consistency, we show that the exact
energy functional for fractional charges, expressed as a linear combination
of the $\gamma_{s}$ of an $N-$electron and that of an $\left(N+1\right)-$electron
excited state, is a straight line interpolating the energies at integers.
We introduce the concepts of excited-state chemical potentials to
describe the slopes of these linear lines. Numerical calculations reveal the
excited-state delocalization error with common approximate functionals
but good performance of corrected functionals on the proven linear
conditions. 
\end{abstract}
\maketitle
Density functional theory (DFT) was originally established as a ground
state theory and has been the major workhorse for electronic structure
calculations in molecules and bulk systems \cite{Hohenberg64B864,Kohn65A1133,Parr89,Dreizler90, Teale2228700}.
The accuracy of DFT calculations critically depends on the density
functional approximations (DFAs) used. Exact conditions for the functional
are critical in functional development \cite{Perdew963865a,Burke12150901,Cohen12289,Becke1418A301,Sun15036402a,Wang202294,Mardirossian16214110}.

In particular, the extension to fractional charges, or electron numbers,
and the exact linear conditions for the energy was originally developed
in a seminal paper by Perdew, Parr, Levy and Baldus based on grand
canonical ensembles \cite{Perdew821691a,Zhang00346}. The PPLB conditions
was later derived based on the exact properties of quantum mechanical
degeneracy and size consistency in the dissociation limits of finite
subsystems \cite{Yang005172}, which further lead to the extensions
to fractional spins and the combination of fractional charges and
spins, and their exact conditions \cite{Cohen08121104,Mori-Sanchez09066403a}.
The ground state PPLB condition plays an important role in understanding
the systematic errors in commonly used DFAs and in developing their
corrections. A major symmetric error in DFAs is the delocalization
error (DE) \cite{Mori-Sanchez08146401}, underlying the underestimation
of bandgaps, reaction barriers, binding energies of charge transfer
complex, and energies of dissociated molecules and the overestimation
of polymer polarizabilities \cite{Cohen08792,Cohen12289,Bryenton22e1631a}.
The DE has been shown to manifest as the convex
deviation from the PPLB linearity conditions for systems small both
in number of atoms and in physical extent, but as the incorrect total
energy differences with the ($N\pm1)$ charged states and consequently
the underestimation of band gaps for bulk systems \cite{Mori-Sanchez08146401}.
To address the challenges, various correction methods have been developed
\cite{Perdew815048,Pederson14121103a,Wagle21094302c,Baer1085,Wing21,Anisimov05075125, Savin96327,Ma1624924,Dabo10115121,Colonna22,Cohen07191109,Zheng11,Li15053001,Li18203,Su201528,Mei22acs.jctc.1c01058, Zhang094963,Su149201b,Li244876}.
Exact conditions for fractional charges and spins have also been developed
for a restricted class of excited states that are the ground states
of some given symmetry and with symmetry-specific functionals of density
(not universal) \cite{Johnson10164107}. 

Deviating from the ground state formulation, DFT has been directly
used for excited-state calculations in the $\Delta SCF$ approach
for a long time\cite{Slater1974AdvQuantumChem,SlaterWood1970SlaterTM},
with much numerical success in predicting excitation energies \cite{ZieglerRaukBaerends1977deltaSCF,GavnholtOlsen2008DeltaSCFes,GilbertBesleyGill2008MOM,BesleyGilbertGill2009DeltaSCF,HarbolaSamal2009esHK,KowalczykanVoorhis2011deltaSCF,MaurerReuter2011deltaSCF,MaurerReuter2013deltaSCF,SeiduZiegler2015constrictedDFT,YeWelbornVanVoorhis2017deltaSCFalgorithm,HaitHeadgordon2020deltaSCF,CarterfenkHerbert2020deltaSCFnumerical,LeviJonsson2020deltaSCF,CorzoPribramjonesHratchian2022PMOM,KumarLuber2022deltaSCF,VandaeleLuber2002deltaSCFreview},
however without a theoretical justification \cite{VandaeleLuber2002deltaSCFreview}, until  recently\cite{Yang24}. To describe excited
states, the electron density alone is not sufficient as the
basic variable. Instead we  use the three sets of equivalent variables
that define the non-interacting reference system: the excitation number
$n_{s}$ and the local one-electron potential $w_{s}({\bf x})$, the
noninteracting wavefunction $\Phi,$ or the 1-particle density matrix
$\gamma_{s}({\bf x},{\bf x}')$. Even though the electron density
is no longer the basic variable, it is still a key quantity in the
theory because the density of the physical system is equal to the
density of the noninteracting reference system for excited states
\cite{Yang24}, as for ground states \cite{Hohenberg64B864,Kohn65A1133,Parr89,Dreizler90}.
It has been shown that ground and excited state energies and densities can be obtained from
the minimum and stationary solutions of the \emph{same} universal
functional \cite{Yang24}. 

We now extend the theory to fractional
charges for excited states,focusing on $\gamma_{s}$ and the density matrix functional
theory ($\gamma_{s}$FT), and following the previous approach for ground
states \cite{Yang005172}. First we consider the simpler case of half integers. Take $N$ electrons
in an external potential $v_{{\bf R}}({\bf r})$ which has a ground
state wavefunction $\Psi_{N}^{0}\left({\bf R}\right)$ with the energy
$E_{v_{{\bf R}}}^{0}(N)$ and the $n$th excited state of $\left(N+1\right)$
electrons in the same external potential $v_{{\bf R}}({\bf r})$ with
a wavefunction $\Psi_{N+1}^{n}\left({\bf R}\right)$ and energy $E_{v_{{\bf R}}}^{n}(N+1)$.
We imply the spin and space coordinates and use ${\bf R}$ to indicate
the location of the external potential. For example, we can just take
${\bf R}$ to be the center-of-mass vector of the nuclei generating
the external potential. The Schroedinger equation for $\Psi_{N+1}^{n}$
is

\begin{equation}
\hat{H}_{{\bf R}}\Psi_{N+1}^{n}\left({\bf R}\right)=E_{v_{{\bf R}}}^{n}(N+1)\Psi_{N+1}^{n}\left({\bf R}\right),\label{phiA-1}
\end{equation}
where the Hamiltonian $\hat{H}_{{\bf R}}$ is composed of the many-electron
kinetic energy, electron-electron interaction energy, and the potential
term from $v_{{\bf R}}({\bf r}).$ Similar equation follows for $\Psi_{N}^{0}\left({\bf R}\right)$.

Now construct a $\left(2N+1\right)$-electron\ system in the external
potential $v({\bf r})=v_{{\bf R}_{1}}({\bf r})+v_{{\bf R}_{2}}({\bf r})$,
with the Hamiltonian $\hat{H}=\hat{H}_{{\bf R}_{1}}+\hat{H}_{{\bf R}_{2}}$
at the limit of $\left|{\bf R}_{1}-{\bf R}_{2}\right|\rightarrow\infty.$
Consider the excited state of this combined system that is simply
composed of an $N$-electron subsystem in its ground state and an
$\left(N+1\right)-$electron subsystems in its excited state $n$,
with two \emph{identical} external potentials separated by an infinite
distance. This state has the wavefunction as the antisymmetric product
of $\Psi_{N}^{0}\left({\bf R}_{1}\right)$-- the ground state at
${\bf R}_{1}$, and $\Psi_{N+1}^{n}\left({\bf R}_{2}\right)$ --the
excited state at ${\bf R}_{2}.$ This state of the combined $\left(2N+1\right)$-electron
system has a two-fold degeneracy: 

\begin{equation}
\Psi_{\alpha}=\hat{A}\left(\Psi_{N}^{0}\left({\bf R}_{1}\right)\Psi_{N+1}^{n}\left({\bf R}_{2}\right)\right),\label{antisym-1}
\end{equation}
and

\begin{equation}
\Psi_{\beta}=\hat{A}\left(\Psi_{N+1}^{n}\left({\bf R}_{1}\right)\Psi_{N}^{0}\left({\bf R}_{2}\right)\right),\label{antisym2-1}
\end{equation}
where $\hat{A}$ is the $\left(2N+1\right)$-electron antisymmetrization
operator. The energy of the degenerate wavefunctions is $E_{v_{{\bf R}}}^{0}(N)+E_{v_{{\bf R}}}^{n}(N+1)$.
Based on the quantum mechanical degeneracy principle, any linear combination
of $\Psi_{\alpha}$ and $\Psi_{\beta}$ is also an eigenstate of our
system with the same energy, including this entangled state:

\begin{equation}
\Psi=\left\{ \Psi_{\alpha}+\Psi_{\beta}\right\} /\sqrt{2}.\label{totalwave-1}
\end{equation}
For the $\Delta SCF$ excited-state theory, from the three sets of equivalent
variable that define the non-interacting reference system \cite{Yang24},
we will focus on the functional of $\gamma_{s},$ which allows the
extension to ensemble states readily, just as the total electron density
for the ground state theory \cite{Mori-Sanchez06201102,Cohen08115123,Ruzsinszky06194112,Vydrov07154109}.
Under the excited-state Kohn-Sham assumption \cite{Yang24}, there
exist the mappings of $\Psi^{0},$$\Psi^{n},$$\Psi_{\alpha}$, $\Psi_{\beta}$
and $\Psi$ to the wavefunctions $\Phi^{0},$$\Phi^{n_{s}},$ $\Phi_{\alpha}$
, $\Phi_{\beta}$ and $\Phi$ of the corresponding noninteracting
reference systems; namely, mapping into $\Psi_{\alpha}$ , $\Psi_{\beta}$
and $\Psi$ are the following:

\begin{equation}
\Phi_{\alpha}=\hat{A}\left(\Phi_{N}^{0}\left({\bf R}_{1}\right)\Phi_{N+1}^{n_{s}}\left({\bf R}_{2}\right)\right),\label{eq:antisym2-1-0}
\end{equation}
\begin{equation}
\Phi_{\beta}=\hat{A}\left(\Phi_{N+1}^{n_{s}}\left({\bf R}_{1}\right)\Phi_{N}^{0}\left({\bf R}_{2}\right)\right),\label{antisym2-1-1}
\end{equation}
\begin{equation}
\Phi=\left\{ \Phi_{\alpha}+\Phi_{\beta}\right\} /\sqrt{2}.\label{eq:totalwave-0}
\end{equation}
The excitation number $n_{s}$ of the noninteracting reference system
does not have to be equal to $n$, the excitation number of the corresponding
interacting system \cite{Yang24}. Furthermore, while $\Psi_{\alpha}$, $\Psi_{\beta},$ and $\Psi$ are $(2N+1)-$electron degenerate wavefunctions
of the \emph{same} many-electron Hamiltonian with the external potential
$v({\bf r})=v_{{\bf R}_{1}}({\bf r})+v_{{\bf R}_{2}}({\bf r})$, the
corresponding noninteracting wavefunctions $\Phi_{\alpha}$ and $\Phi_{\beta}$
have the same noninteracting eigenvalues, but are \emph{not} degenerate
wavefunctions of a single noninteracting reference Hamiltonian $H_{s}$.
Instead, they are the eigenstates of two different noninteracting
reference Hamiltonians: $\Phi_{\alpha}$ is an eigenstate of a $H_{s}$
with potential $w_{s,{\bf R}_{1}}^{0}({\bf x},N)+w_{s,{\bf R}_{2}}^{n_{s}}({\bf x},N+1)$,
and $\Phi_{\beta}$ is an eigenstate of another $H_{s}$ with potential
$w_{s,{\bf R}_{1}}^{n_{s}}({\bf x},N+1)+w_{s,{\bf R}_{2}}^{0}({\bf x},N)$,
where we use the notation $w_{s,{\bf R}_{2}}^{n_{s}}({\bf x},N)$
to denote the local potential associated with the $n_{s}$th eigenstate
$\Phi_{N+1}^{n_{s}}\left({\bf R}_{2}\right).$ Therefore, the two-fold
degeneracy for the interacting Hamiltonian is represented as two different
noninteracting wavefunctions and Hamiltonians, with equal total noninteracting
energies. It is highly interesting that the symmetry of the interacting
Hamiltonian is broken here by the corresponding noninteracting Hamiltonians
\cite{Su189678a,Perdew21e2017850118b}.

The one-electron density matrix of $\Phi$ is 

\begin{align}
\gamma_{s}({\bf x},{\bf x}') & =(\gamma_{s{\bf R}_{1}}^{0}(N)+\gamma_{s{\bf R}_{1}}^{n_{s}}(N+1)+\gamma_{s{\bf R}_{2}}^{0}(N)\nonumber \\
 & +\gamma_{s{\bf R}_{2}}^{n_{s}}(N+1))/2+C_{1}({\bf x},{\bf x}'),\label{E2E0-2}
\end{align}
where $\gamma_{s{\bf R}_{1}}^{0}(N)/\gamma_{s{\bf R}_{1}}^{n_{s}}(N+1)$,
with the coordinates $({\bf x},{\bf x}')$ implied, denotes the density
matrix of $\Phi_{N}^{0}\left({\bf R}_{1}\right)/\Phi_{N+1}^{n_{s}}\left({\bf R}_{1}\right)$.
$C_{1}({\bf x},{\bf x}')$ is the contribution to $\gamma_{s}({\bf x},{\bf x}')$
from the cross terms like $\left\langle \Phi_{\alpha}\left|\sum_{i}\delta\left({\bf x}-{\bf x}_{i}\right)\delta\left({\bf x}'-{\bf x}'_{i}\right)\right|\Phi_{\beta}\right\rangle $.
Importantly, $C_{1}({\bf x},{\bf x}')=0$, for all $\left|{\bf r}-{\bf r}'\right|<\infty,$
because $\Phi^{0}$ and $\Phi^{n_{s}}$ have different particle numbers
$N$ and $N-1$, and any coordinate integration of the overlap of
$\Phi^{0}$ at ${\bf R}_{1}$ with $\Phi^{n_{s}}$ at ${\bf R}_{2}$
is zero (in other words, $\Phi_{N}^{0}\left({\bf R}_{1}\right)$ and
$\Phi_{N+1}^{n_{s}}\left({\bf R}_{2}\right)$ are strongly orthogonal
\cite{Arai6095,Herbert07261a}). Details are provided in the SI.

Since $C_{1}({\bf x},{\bf x}')=0$ for all ${\bf r}$ and ${\bf r}'$
except at $\left|{\bf r}-{\bf r}'\right|\longrightarrow\infty$, it
cannot contribute to the energy of the system (see SI and Ref.\cite{Nakata12032125a}
). Therefore we have $E_{v}\left[\gamma_{s}({\bf x},{\bf x}')\right]$=$E_{v}[(\gamma_{s{\bf R}_{1}}^{0}+\gamma_{s{\bf R}_{1}}^{n_{s}}+\gamma_{s{\bf R}_{2}}^{0}+\gamma_{s{\bf R}_{2}}^{n_{s}})/2]$.
Now we consider the behavior of the energy functional, $E_{v}\left[\gamma_{s}({\bf x},{\bf x}')\right]$.
We do not restrict ourselves to any specific formulation of the energy
functional, instead assuming certain exact properties; namely:

1)$E_{v}\left[\gamma_{s}\right]$ gives the correct energy. \ Hence,
for the total density matrix in Eq. (\ref{E2E0-2}), we have: 
\begin{align}
 & E_{v}[(\gamma_{s{\bf R}_{1}}^{0}+\gamma_{s{\bf R}_{1}}^{n_{s}}+\gamma_{s{\bf R}_{2}}^{0}+\gamma_{s{\bf R}_{2}}^{n_{s}})/2]\nonumber \\
 & =E_{v_{{\bf R}_{1}}}^{0}(N)+E_{v_{{\bf R}_{2}}}^{n}(N+1).\label{E2E0}
\end{align}

2)$E_{v}\left[\gamma_{s}\right]$ is size consistent. \ Therefore,

\begin{align}
 & E_{v}[(\gamma_{s{\bf R}_{1}}^{0}+\gamma_{s{\bf R}_{1}}^{n_{s}}+\gamma_{s{\bf R}_{2}}^{0}+\gamma_{s{\bf R}_{2}}^{n_{s}})/2]\nonumber \\
 & =E_{v_{{\bf R}_{1}}}[(\gamma_{s{\bf R}_{1}}^{0}+\gamma_{s{\bf R}_{1}}^{n_{s}})/2]+E_{v_{{\bf R}_{2}}}[(\gamma_{s{\bf R}_{2}}^{0}+\gamma_{s{\bf R}_{2}}^{n_{s}})/2].\label{size-consistent}
\end{align}.

3)$E_{v}\left[\gamma_{s}\right]$ is translationally invariant. \ Therefore,
\begin{equation}
E_{v_{{\bf R}_{1}}}[(\gamma_{s{\bf R}_{1}}^{0}+\gamma_{s{\bf R}_{1}}^{n_{s}})/2]=E_{v_{{\bf R}_{2}}}[(\gamma_{s{\bf R}_{2}}^{0}+\gamma_{s{\bf R}_{2}}^{n_{s}})/2].\label{transinvariant}
\end{equation}
From Eq. (\ref{transinvariant}) and Eq. (\ref{E2E0}) it follows
that

\begin{align}
 & E_{v_{{\bf R}_{1}}}[(\gamma_{s{\bf R}_{1}}^{0}(N)+\gamma_{s{\bf R}_{1}}^{n_{s}}(N+1))/2]\nonumber \\
 & =\frac{1}{2}(E_{v_{{\bf R}_{1}}}^{0}(N)+E_{v_{{\bf R}_{2}}}^{n}(N+1)).\label{E2degenerate}
\end{align}

This shows that for {\em any} energy functional, which satisfies
the exact properties of degeneracy, size-extensivity, and translational
invariance, its value for the density matrix $(\gamma_{s{\bf R}_{1}}^{0}(N)+\gamma_{s{\bf R}_{1}}^{n_{s}}(N+1))/2$
is just the linear interpolation of corresponding energies at integers. 

The ground-state nature of \textbf{$\Phi^{0}$} plays no particular
role in the derivation and the foregoing results can be derived equally
for any two states\textbf{ }with excitation numbers $n$ and $m$
for the physical system and with excitation numbers $n_{s}$ and $m_{s}$
for the corresponding noninteracting reference systems. Thus we can
obtain the general excited-state fractional charge conditions for
half integers

\begin{align}
 & E_{v_{{\bf R}_{1}}}[\left(\gamma_{s{\bf R}_{1}}^{n_{s}}(N)+\gamma_{s{\bf R}_{1}}^{m_{s}}(N+1)\right)/2]\nonumber \\
 & =\frac{1}{2}\left(E_{v_{{\bf R}_{1}}}^{n}(N)+E_{v_{{\bf R}_{1}}}^{m}(N+1)\right).\label{E2degenerate-1}
\end{align}

We now extend the result of Eq. (\ref{E2degenerate-1}) to a general
combination of the two density matrices. Consider a system with $qN+p$
electrons, where $p$, $q(>p)$ and $N$ are integers, in the external
potential$\ v({\bf r})=\sum_{i=1}^{q}v_{{\bf R}_{i}}({\bf r})$ at
the limit of $\left|{\bf R}_{i}-{\bf R}_{j}\right|\rightarrow\infty,$
$\in i,j.$ Then the total system is simply composed of $q$ systems
with identical external potential $v_{{\bf R}_{i}}({\bf r})$ separated
by infinite distances. Consider an excited state of this system, among
the $q$ systems, $p$ systems have $N+1$ electrons with an excited
state wavefunction $\Psi_{N+1}^{m}\left(\left\{ {\bf x}\right\} \right)$
and remaining $(q-p)$ systems to have $N$ electrons with the $n$th
excited state wavefunction $\Psi_{N}^{n}\left(\left\{ {\bf x}\right\} \right)$,
both under the same potential at different locations. The total ground-state
wavefunction is the antisymmetric product of $q$ separated wavefunctions,
with a total energy of $(q-p)E_{v_{{\bf R}_{1}}}^{n}(N)+pE_{v_{{\bf R}_{1}}}^{m}(N+1).$
\ Among the possible degenerate wavefunctions, one is a state $\Psi_{1}$,
in which the first $p$ locations, ${\bf R}_{1}...{\bf R}_{p},$ have
the wavefunction $\Psi_{N+1}^{m}$ and the rest the other wavefunction
$\Psi_{N}^{n};$ namely,

\begin{eqnarray}
\Psi_{1} & = & \hat{A}\{\Psi_{N+1}^{m}\left({\bf R}_{1}\right)\Psi_{N+1}^{m}\left({\bf R}_{2}\right)...\Psi_{N+1}^{m}\left({\bf R}_{p}\right)\nonumber \\
 &  & \Psi_{N}^{n}\left({\bf R}_{p+1}\right)\Psi_{N}^{n}\left({\bf R}_{p+2}\right)...\Psi_{N}^{n}\left({\bf R}_{q}\right)\},\label{antisym4}
\end{eqnarray}
where the space and spin coordinates of $N$ or $(N+1)$ electrons
are implied. Permutation of any two locations with $\Psi_{N+1}^{m}$
and $\Psi_{N}^{n}$ generates a different degenerate wavefunction.
There are a total of $M=\frac{q!}{p!(q-p)!}$ such degenerate wavefunctions,
$\{\Psi_{1},\Psi_{2},...\Psi_{M}\}$. \ For any wave function $\Psi_{k},$
a particular location ${\bf R}_{p}$ can either have the wavefunction
$\Psi_{N}^{n}$ or $\Psi_{N+1}^{m}$. As in Eq.(\ref{eq:totalwave-0}),
the maximally entangled state $\Psi=\frac{1}{\sqrt{M}}\sum_{k=1}^{M}\Psi_{k}$
is also a degenerate wavefunction. In all the product state wavefunctions
$\{\Psi_{1},\Psi_{2}...\Psi_{M}\}$, the number of times any location
${\bf R}_{p},$having the wave function $\Psi_{N+1}^{m}$ is equal
to $M_{N+1}=\frac{(q-1)!}{(p-1)!(q-p)!},$ and the corresponding number
for $\Psi_{N}^{n}$ is $M_{N}=M-M_{N+1}$. Now in the $\Delta SCF$
 theory, $\Psi_{N}^{n}$ and $\Psi_{N+1}^{m}$ correspond to their
noninteracting wavefunction $\Phi_{N}^{n_{s}}$ and $\Phi_{N+1}^{m_{s}}$.
Thus, the many-electron product states $\{\Psi_{1},\Psi_{2},...\Psi_{M}\}$
correspond to the noninteracting reference wavefunctions that are
the product states$\{\Phi_{1},\Phi_{2},...\Phi_{M}\}$. 

Corresponding to the maximally entangled many-electron state $\Psi=\frac{1}{\sqrt{M}}\sum_{k=1}^{M}\Psi_{k}$, the maximally entangled noninteracting reference wavefunction is
$\Phi=\frac{1}{\sqrt{M}}\sum_{k=1}^{M}\Phi_{k}$. Its density matrix is
\begin{eqnarray}
\gamma_{s} & = & \frac{1}{M}\sum_{t=1}^{q}(M_{N+1}\gamma_{s{\bf {\bf R}}_{t}}^{m_{s}}(N+1)\nonumber \\
 &  & +(M-M_{N+1})\gamma_{s{\bf R}_{t}}^{n_{s}}(N))+C_{2}\nonumber \\
 & = & \sum_{t=1}^{q}\frac{p}{q}\gamma_{s{\bf {\bf R}}_{t}}^{m_{s}}(N+1)+\frac{q-p}{q}\gamma_{s{\bf R}_{t}}^{n_{s}}(N)+C_{2},\label{totalRho3}
\end{eqnarray}
where $C_{2}({\bf x},{\bf x}')$ is the contributions from cross terms
like $\left\langle \Phi_{k}\left|\sum_{i}\delta\left({\bf x}-{\bf x}_{i}\right)\delta\left({\bf x}'-{\bf x}'_{i}\right)\right|\Phi_{l}\right\rangle $
and dose not contribute to the energy functional, as $C_{1}({\bf x},{\bf x}')$
in Eq. (\ref{E2E0-2}). Following the arguments of Eqs.(\ref{E2E0}-\ref{E2degenerate}),
we obtain the general linearity conditions for fractional charges
in excited states:

\begin{align}
 & E_{v}[\frac{q-p}{q}\gamma_{s}^{n_{s}}(N)+\frac{p}{q}\gamma_{s}^{m_{s}}(N+1)]\nonumber \\
= & \frac{q-p}{q}E_{v}^{n}(N)+\frac{p}{q}E_{v}^{m}(N+1),\label{GeneralLinearity}
\end{align}
where we have dropped the reference to the location ${\bf R}_{p}.$
This is our key result. It agrees with PPLB linear conditions on fractional
charges in the special case of ground states:$n_{s}=m_{s}=n=m=0,$and
when the basic variable are the electron densities, the diagonal elements
of $\gamma_{s}$. It extends the PPLB linear conditions in two
key aspects: the basic variables for the fractional charge systems are
now $\gamma_{s},$ the 1-particle density matrices of the noninteracting
reference systems, and the $N-$ and $\left(N+1\right)-$electron
states are all states, including the ground states. 

For ground states, the PPLB linear conditions set the physical meaning
for ground-state chemical potentials\cite{Parr783801}, $\mu=\left(\frac{\partial E_{v}(N)}{\partial N}\right)_{v}$,
the slopes of $E_{v}(N)$: $\mu=-I$ for $N-\delta$, and $\mu=-A$
for $N+\delta$, where the ionization energy $I=E_{v}(N-1)-E_{v}(N)$
and the electron affinity $A=E_{v}(N)-E_{v}(N+1)$ \cite{Perdew821691a,Parr89}.

For excited states, the slopes of the linear curves in Eq. (\ref{GeneralLinearity})
also convey physical meanings. For a given $n_{s}$th eigenstate with
an integer particle number $N,$ on the electron addition side, the
fractional electron number $\mathit{\mathcal{N}}$ connecting to the
$m_{s}$th eigenstate of the $(N+1)-$electron system (described by
$\gamma_{s}^{m_{s}}(N+1)$) is $\mathit{\mathcal{N}}=\frac{q-p}{q}N+\frac{p}{q}(N+1)$, 
and the energy as a function of  $\mathit{\mathcal{N}}$ is $E_{v}^{+}(n_{s},m_{s},\mathit{\mathcal{N}})=E_{v}[\frac{q-p}{q}\gamma_{s}^{n_{s}}(N)+\frac{p}{q}\gamma_{s}^{m_{s}}(N+1)],$
the left hand side of Eq. (\ref{GeneralLinearity}). We define the
\emph{excited-state chemical potentials} $\mu_{n_{s}m_{s}}^{+}$as
the slopes of $E_{v}^{+}(n_{s},m_{s},\mathit{\mathcal{N}})$. For
$N<\mathcal{N}<N+1$, $\mu_{n_{s}m_{s}}^{+}$ is a constant:\\
\begin{equation}
\mu_{n_{s}m_{s}}^{+}=\left(\frac{\partial E_{v}^{+}(n_{s},m_{s},\mathit{\mathcal{N}})}{\partial\mathit{\mathcal{N}}}\right)_{v}=E_{v}^{m}(N+1)-E_{v}^{n}(N),\label{eq:mu_plus}
\end{equation}
where we used the right hand side of Eq. (\ref{GeneralLinearity}). 

Similarly on the electron removal side, the equivalent of Eq. (\ref{GeneralLinearity})
is

\begin{align}
 & E_{v}[\frac{q-p}{q}\gamma_{s}^{n_{s}}(N)+\frac{p}{q}\gamma_{s}^{l_{s}}(N-1)]\nonumber \\
= & \frac{q-p}{q}E_{v}^{n}(N)+\frac{p}{q}E_{v}^{l}(N-1).\label{eq:GeneralLinearity_minus}
\end{align}
The fractional electron number $\mathit{\mathcal{N}}$ connecting to
the $l_{s}$th eigenstate of the $(N-1)-$electron system (described
by $\gamma_{s}^{l_{s}}(N-1)$) is $\mathit{\mathcal{N}}=\frac{q-p}{q}N+\frac{p}{q}(N-1)$,
and the energy as a function of $\mathit{\mathcal{N}}$ is $E_{v}^{-}(n_{s},l_{s},\mathit{\mathcal{N}})=E_{v}[\frac{q-p}{q}\gamma_{s}^{n_{s}}(N)+\frac{p}{q}\gamma_{s}^{l_{s}}(N-1)].$
We then define the excited-state chemical potentials $\mu_{n_{s}m_{s}}^{-}$as
the slopes of $E_{v}^{-}(n_{s},l_{s},\mathit{\mathcal{N}})$ . For
$N-1<\mathcal{N}<N$,

\begin{equation}
\mu_{n_{s}l_{s}}^{-}=\left(\frac{\partial E_{v}^{-}(n_{s},l_{s},\mathit{\mathcal{N}})}{\partial\mathit{\mathcal{N}}}\right)_{v}=E_{v}^{n}(N)-E_{v}^{l}(N-1),\label{eq:mu_minus}
\end{equation}
where we used the right hand side of Eq. (\ref{eq:GeneralLinearity_minus}).
There is an symmetry: $\mu_{n_{s}m_{s}}^{+}(N)$=$\mu_{m_{s}n_{s}}^{-}(N+1)$.
For ground states, $\mu_{00}^{+}$ and $\mu_{00}^{-}$ agree with
the chemical potentials of derived from the PPLB condition \cite{Perdew821691a}.
The excited state chemical potentials are the negative of IP of an excited state as in Eq. (\ref{eq:mu_minus}), or the
negative of EA of an excited state as in Eq.(\ref{eq:mu_plus}).
The excited state chemical potentials should play important roles
in $\Delta SCF$ theory, as they do in the ground state
DFT.
\begin{figure}[h]
\centering{}\includegraphics[width=3.4in]{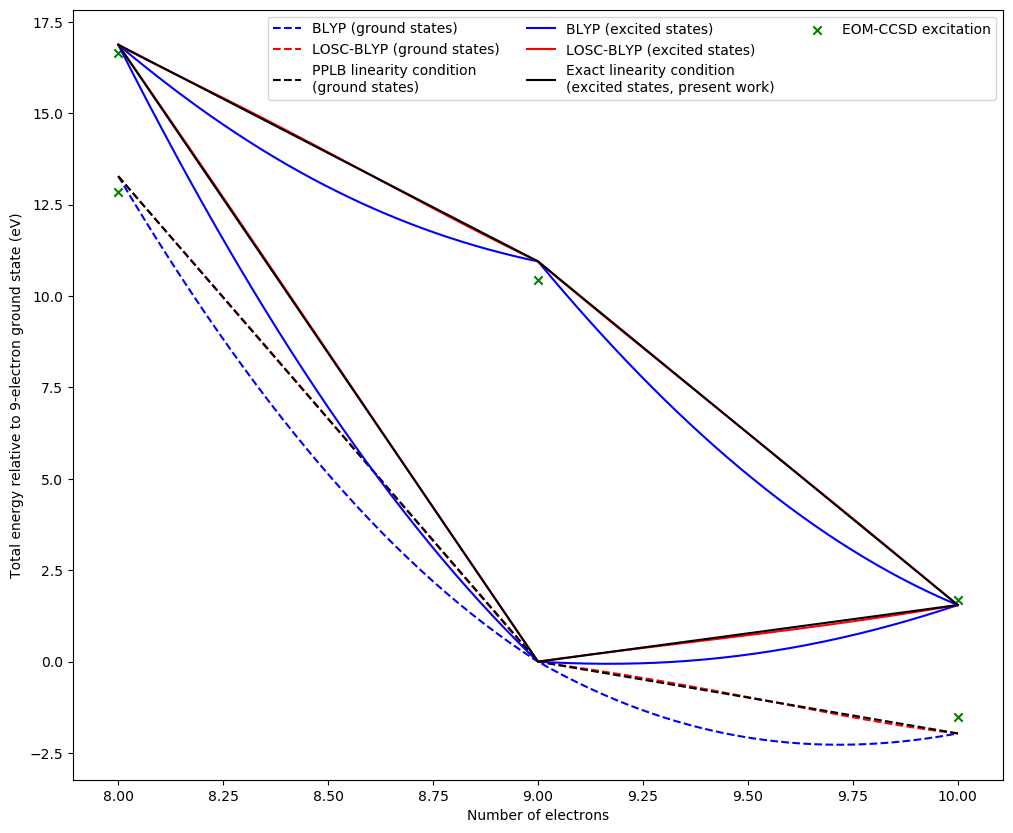}\caption{The total energies of fractional charges from the ground and the ($\sigma\rightarrow\sigma^{*})$
excited state of the $\text{OH}$ molecule $(N=9)$ to the ground and the
($n\rightarrow\sigma^{*})$ excited state of $\text{OH}^{-}$ $(N=10)$ and
also to the ground and the ($\sigma\rightarrow n)$ excited state
of $\text{OH}^{+}$ $(N=8)$. All the relative energies at integer electron
numbers from the BLYP calculations well approximate the results from
EOM-CCSD calculations in green cross (details in SI). Note that  LOSC-BLYP  gives numerically the same total energies as BLYP  for small systems ( $\text{OH}$, $\text{OH}^{+}$ and $\text{OH}^{-}$) by design in LOSC.   For the ground
states, the exact PPLB conditions for fractional charges (two lowest-energy
dash black lines) are significantly underestimated by the BLYP results,
but are well approximated by the LOSC-BLYP\cite{Mei2021,yu_accurate_2024,williams_correcting_2024}
results. The exact conditions for excited-state fractional charges,
proved presently, are represented by the four solid black lines. These
four exact lines are all significantly underestimated by the BLYP
results, but are well approximated by the LOSC-BLYP results, as in
ground states.}
\end{figure}

The present introduction of fractional charges and excited-state chemical
potentials also allows us to explore the chemical concepts for excited
states, leading to excited state electronegativity, hardness, and
fukui functions, extending corresponding ground-state concepts \cite{Geerlings031793,Parr783801,Parr837512,Parr844049a}
(See SI for details). These excited-state chemical concepts can be
useful for describing chemical reactivity in excited states. 

Finally, we examine the performance of commonly used DFAs on the newly
derived exact conditions, Eq. (\ref{GeneralLinearity}). Numerical
results in Fig 1 lead us to define the \emph{excited-state delocalization
error} (exDE), similarly to DE in ground states \cite{Mori-Sanchez08146401}
in commonly used DFAs.  More details and additional results are presented 
in SI. Our calculations shows that for small systems with commonly used DFAs, the  exDE is demonstrated as the convex deviation for fractional charges from the exact conditions presently proved, however the excitation energies for physical systems with integer changes are  well approximated by the $\Delta$SCF total energy calculations. The recently developed LOSC method
for correcting the DE for ground state calculations \cite{Li18203,Su2020,Mei2021,yu_accurate_2024,williams_correcting_2024}
shows excellent agreement with the excited-state linearity conditions.
We expect that the excited-state delocalization error in commonly used DFAs would have similar size-dependent manifestation as in ground states \cite{Mori-Sanchez08146401}:
when a system gets larger and approaches the bulk limit, the observed convex deviation  from the exact linear conditions would decrease and disappear, but the error is shifted to the excited-state  energy differences from the $\Delta$SCF total energy calculations.

In summary, we have the extended $\Delta SCF$ excited-state theory to
systems with fractional charges and derived the linearity conditions
for the functional of the noninteracting reference density matrix
based on the exact properties of degeneracy and size consistency. We defined the slopes of the linear lines
connecting all states as the excited-state chemical potentials, and 
they convey physical meaning of the corresponding excited-state IPs and EAs. The linear conditions derived are exact conditions
on the density matrix functional $E_{v}[\gamma_{s}]$ for ground
state and excited states. They allow us to introduce presently the concept of excited-state delocalization error to understand the observed deviation from commonly used DFAs.  They should play important roles for
understanding and designing functional approximations for all states.

We acknowledge support from the National Science Foundation (CHE-2154831)
and the National Institute of Health (R01-GM061870).

\bibliography{FractionalDeltaSCF_2024Notes,FractionalDeltaSCF_2024_add}

\end{document}